\batchmode
\makeatletter
\def\input@path{{"/home/jacob/Documents/Work/My Papers/2026-PVM-POVM Theorems/"}}
\makeatother
\documentclass[11pt,twoside,english]{article}
\usepackage[T1]{fontenc}
\usepackage[latin9]{inputenc}
\usepackage{color}
\definecolor{lyxboxbgcolor}{rgb}{0.980469, 0.941406, 0.902344}
\usepackage{babel}
\usepackage{amsmath}
\usepackage{amssymb}
\usepackage{geometry}
\usepackage{setspace}
\usepackage{xargs}[2008/03/08]
\usepackage[pdftex,pdfusetitle,
 bookmarks=true,bookmarksnumbered=true,bookmarksopen=false,
 breaklinks=false,pdfborder={0 0 0},pdfborderstyle={},backref=false,colorlinks=false]
 {hyperref}

\makeatletter
\let\originalleft\left
\let\originalright\right
\renewcommand{\left}{\mathopen{}\mathclose\bgroup\originalleft}
\renewcommand{\right}{\aftergroup\egroup\originalright}

\def\smalloverbrace#1{\mathop{\vbox{\m@th\ialign{##\crcr%
      \noalign{\kern3\p@}%
      \tiny\downbracefill\crcr\noalign{\kern3\p@\nointerlineskip}%
      $\hfil\displaystyle{#1}\hfil$\crcr}}}\limits}

\def\smallunderbrace#1{\mathop{\vtop{\m@th\ialign{##\crcr
   $\hfil\displaystyle{#1}\hfil$\crcr
   \noalign{\kern3\p@\nointerlineskip}%
   \tiny\upbracefill\crcr\noalign{\kern3\p@}}}}\limits}

\makeatother

\begin{document}
\title{The Born Representation Theorem\\
and the Unistochastic Theorem}
\author{Jacob A. Barandes\thanks{Departments of Philosophy and Physics, Harvard University, Cambridge, MA 02138; jacob\_barandes@harvard.edu; ORCID: 0000-0002-3740-4418}
}
\date{}

\maketitle

\begin{abstract}
This paper presents self-contained, constructive proofs of two new
theorems about stochastic matrices, with direct relevance to quantum
theory. The first theorem, herein called the Born Representation Theorem,
shows that each entry of any stochastic matrix can be expressed as
the trace of a pairwise product of matrices, where the first factor
in the pairwise product belongs to a positive-operator-valued measure
(POVM) and the second factor belongs to a projection-valued measure
(PVM). As its name suggests, this theorem entails that the entries
of any stochastic matrix can be expressed in terms of a generalized
version of the quantum-theoretic Born rule. It follows as a corollary
that if the POVM in this first theorem is a PVM, then the stochastic
matrix is unistochastic, meaning that its entries are each the modulus
square of the corresponding entry of a unitary matrix of the same
size. The second theorem proved in this paper, called the Unistochastic
Theorem, then shows that by dilating the underlying vector space by
a bounded number of additional dimensions if necessary, each entry
of any stochastic matrix can be expressed in terms of the trace of
a pairwise product for which both factors belong to PVMs, and can
thus be derived via marginalization from a larger unistochastic matrix.
This second theorem therefore establishes a kind of primacy of unistochastic
matrices over stochastic matrices, and hints at a close connection
with unitary time evolution in quantum theory. The paper concludes
with a brief discussion of potential applications to discrete-time
deterministic processes and Markov chains.
\end{abstract}

\begin{center}
\global\long\def\quote#1{``#1"}%
\global\long\def\apostrophe{\textrm{'}}%
\global\long\def\slot{\phantom{x}}%
\global\long\def\eval#1{\left.#1\right\vert }%
\global\long\def\keyeq#1{\boxed{#1}}%
\global\long\def\importanteq#1{\boxed{\boxed{#1}}}%
\global\long\def\given{\vert}%
\global\long\def\mapping#1#2#3{#1:#2\to#3}%
\global\long\def\composition{\circ}%
\global\long\def\set#1{\left\{  #1\right\}  }%
\global\long\def\setindexed#1#2{\left\{  #1\right\}  _{#2}}%

\global\long\def\setbuild#1#2{\left\{  \left.\!#1\,\right|\,#2\right\}  }%
\global\long\def\complem{\mathrm{c}}%

\global\long\def\union{\cup}%
\global\long\def\intersection{\cap}%
\global\long\def\cartesianprod{\times}%
\global\long\def\disjointunion{\sqcup}%

\global\long\def\isomorphic{\cong}%

\global\long\def\setsize#1{\left|#1\right|}%
\global\long\def\defeq{\equiv}%
\global\long\def\conj{\ast}%
\global\long\def\overconj#1{\overline{#1}}%
\global\long\def\re{\mathrm{Re\,}}%
\global\long\def\im{\mathrm{Im\,}}%

\global\long\def\transp{\mathrm{T}}%
\global\long\def\tr{\mathrm{tr}}%
\global\long\def\adj{\dagger}%
\global\long\def\diag#1{\mathrm{diag}\left(#1\right)}%
\global\long\def\dotprod{\cdot}%
\global\long\def\crossprod{\times}%
\global\long\def\Probability#1{\mathrm{Prob}\left(#1\right)}%
\global\long\def\Amplitude#1{\mathrm{Amp}\left(#1\right)}%
\global\long\def\cov{\mathrm{cov}}%
\global\long\def\corr{\mathrm{corr}}%

\global\long\def\absval#1{\left\vert #1\right\vert }%
\global\long\def\expectval#1{\left\langle #1\right\rangle }%
\global\long\def\op#1{\hat{#1}}%

\global\long\def\bra#1{\left\langle #1\right|}%
\global\long\def\ket#1{\left|#1\right\rangle }%
\global\long\def\braket#1#2{\left\langle \left.\!#1\right|#2\right\rangle }%

\global\long\def\parens#1{(#1)}%
\global\long\def\bigparens#1{\big(#1\big)}%
\global\long\def\Bigparens#1{\Big(#1\Big)}%
\global\long\def\biggparens#1{\bigg(#1\bigg)}%
\global\long\def\Biggparens#1{\Bigg(#1\Bigg)}%
\global\long\def\bracks#1{[#1]}%
\global\long\def\bigbracks#1{\big[#1\big]}%
\global\long\def\Bigbracks#1{\Big[#1\Big]}%
\global\long\def\biggbracks#1{\bigg[#1\bigg]}%
\global\long\def\Biggbracks#1{\Bigg[#1\Bigg]}%
\global\long\def\curlies#1{\{#1\}}%
\global\long\def\bigcurlies#1{\big\{#1\big\}}%
\global\long\def\Bigcurlies#1{\Big\{#1\Big\}}%
\global\long\def\biggcurlies#1{\bigg\{#1\bigg\}}%
\global\long\def\Biggcurlies#1{\Bigg\{#1\Bigg\}}%
\global\long\def\verts#1{\vert#1\vert}%
\global\long\def\bigverts#1{\big\vert#1\big\vert}%
\global\long\def\Bigverts#1{\Big\vert#1\Big\vert}%
\global\long\def\biggverts#1{\bigg\vert#1\bigg\vert}%
\global\long\def\Biggverts#1{\Bigg\vert#1\Bigg\vert}%
\global\long\def\Verts#1{\Vert#1\Vert}%
\global\long\def\bigVerts#1{\big\Vert#1\big\Vert}%
\global\long\def\BigVerts#1{\Big\Vert#1\Big\Vert}%
\global\long\def\biggVerts#1{\bigg\Vert#1\bigg\Vert}%
\global\long\def\BiggVerts#1{\Bigg\Vert#1\Bigg\Vert}%
\global\long\def\ket#1{\vert#1\rangle}%
\global\long\def\bigket#1{\big\vert#1\big\rangle}%
\global\long\def\Bigket#1{\Big\vert#1\Big\rangle}%
\global\long\def\biggket#1{\bigg\vert#1\bigg\rangle}%
\global\long\def\Biggket#1{\Bigg\vert#1\Bigg\rangle}%
\global\long\def\bra#1{\langle#1\vert}%
\global\long\def\bigbra#1{\big\langle#1\big\vert}%
\global\long\def\Bigbra#1{\Big\langle#1\Big\vert}%
\global\long\def\biggbra#1{\bigg\langle#1\bigg\vert}%
\global\long\def\Biggbra#1{\Bigg\langle#1\Bigg\vert}%
\global\long\def\braket#1#2{\langle#1\vert#2\rangle}%
\global\long\def\bigbraket#1#2{\big\langle#1\big\vert#2\big\rangle}%
\global\long\def\Bigbraket#1#2{\Big\langle#1\Big\vert#2\Big\rangle}%
\global\long\def\biggbraket#1#2{\bigg\langle#1\bigg\vert#2\bigg\rangle}%
\global\long\def\Biggbraket#1#2{\Bigg\langle#1\Bigg\vert#2\Bigg\rangle}%
\global\long\def\angs#1{\langle#1\rangle}%
\global\long\def\bigangs#1{\big\langle#1\big\rangle}%
\global\long\def\Bigangs#1{\Big\langle#1\Big\rangle}%
\global\long\def\biggangs#1{\bigg\langle#1\bigg\rangle}%
\global\long\def\Biggangs#1{\Bigg\langle#1\Bigg\rangle}%

\global\long\def\vec#1{\mathbf{#1}}%
\global\long\def\vecgreek#1{\boldsymbol{#1}}%
\global\long\def\idmatrix{1\!\!1}%
\global\long\def\projector{P}%
\global\long\def\permutationmatrix{\Sigma}%
\global\long\def\densitymatrix{\rho}%
\global\long\def\krausmatrix{K}%
\global\long\def\stochasticmatrix{\Gamma}%
\global\long\def\lindbladmatrix{L}%
\global\long\def\dynop{\Theta}%
\global\long\def\timeevop{U}%
\global\long\def\hadamardprod{\odot}%
\global\long\def\tensorprod{\otimes}%

\global\long\def\inprod#1#2{\left\langle #1,#2\right\rangle }%
\global\long\def\normket#1{\left\Vert #1\right\Vert }%
\global\long\def\hilbspace{\mathcal{H}}%
\global\long\def\samplespace{\Omega}%
\global\long\def\configspace{\mathcal{C}}%
\global\long\def\phasespace{\mathcal{P}}%
\global\long\def\spectrum{\sigma}%
\global\long\def\restrict#1#2{\left.#1\right\vert _{#2}}%
\global\long\def\from{\leftarrow}%
\global\long\def\statemap{\omega}%
\global\long\def\degangle#1{#1^{\circ}}%
\global\long\def\trivialvector{\tilde{v}}%
\global\long\def\eqsbrace#1{\left.#1\qquad\right\}  }%
\global\long\def\operator#1{\operatorname{#1}}%
\global\long\def\liff{\leftrightarrow}%
\global\long\def\liffwide{\ \leftrightarrow\ }%
\global\long\def\limplies{\rightarrow}%
\global\long\def\lxor{\nleftrightarrow}%
\global\long\def\lproposition{P}%
\global\long\def\ltrue{T}%
\global\long\def\lfalse{F}%
\global\long\def\taut{\mathbb{T}}%
\global\long\def\antitaut{\mathbb{F}}%
\global\long\def\probvariable{\rho}%
\newcommandx\cptpmap[2][usedefault, addprefix=\global, 1=, 2=]{\mathcal{E}^{#2}_{#1}}%
\newcommandx\auxstochasticmap[2][usedefault, addprefix=\global, 1=, 2=]{\mathcal{F}^{#2}_{#1}}%
\global\long\def\matf#1{\mathbf{#1}}%
\global\long\def\matvec#1{\mathbf{#1}}%
\global\long\def\repmap{\pi}%
\global\long\def\densitymatrix{\boldsymbol{\rho}}%
\global\long\def\matham{\mathbf{H}}%
\global\long\def\matgreek#1{\boldsymbol{#1}}%
\global\long\def\clalgebra{\mathcal{A}}%
\global\long\def\qualgebra{\mathcal{Q}}%
\global\long\def\idmatrix{1\!\!1}%
\global\long\def\idvar{1\!\!1}%
\global\long\def\zerovar{0\!\!0}%
\global\long\def\zeroket{0}%
\par\end{center}

\section{Introduction\label{sec:Introduction}}

This paper presents and proves two new theorems as part of an exploration
into the deep connections between stochastic matrices, such as those
used in modeling discrete stochastic processes, and the probability
formulas of quantum theory.

An $M\times N$ stochastic matrix $\matgreek{\stochasticmatrix}$
is a matrix whose entries are all non-negative real numbers $\stochasticmatrix_{ij}\geq0$
and each of whose columns sums to unity: $\sum^{M}_{i=1}\stochasticmatrix_{ij}=1$.
In some references, such a matrix is said to be column stochastic
or left stochastic. One can use a stochastic matrix to model a conditional
probability distribution $p\left(i\given j\right)\equiv p\left(i\textrm{ given }j\right)$
over a finite set of states or configurations indexed by $i=1,\dots,M$
and $j=1,\dots,N$, with $p\left(i\given j\right)\defeq\stochasticmatrix_{ij}$.

The first theorem proved in this paper shows that any stochastic matrix
$\matf{\stochasticmatrix}$ can be expressed, entry by entry, as a
matrix trace, $\stochasticmatrix_{ij}=\tr\left(\matf E_{i}\matf{\projector}_{j}\right)$,
where ${\{\matf E_{i}\}}_{i=1,\dots,M}$ is an $M$-member collection
of $N\times N$ matrices forming a positive-operator-valued measure
(POVM) and where ${\{\matf{\projector}_{j}\}}_{j=1,\dots,N}$ is an
$N$-member collection of $N\times N$ matrices forming a projection-valued
measure (PVM). This trace formula precisely coincides with the quantum-theoretic
Born rule (originally due to Born 1926)\nocite{Born:1926zqds} for
the conditional probability $p\left(i\given j\right)$ with which
a quantum system described by a pure state, represented by $\matf{\projector}_{j}$
for some $j=1,\dots,N$, will yield a specific result or effect, represented
by $\matf E_{i}$ for some $i=1,\dots,M$, following a generalized
measurement. This first theorem will therefore be called the Born
Representation Theorem.

A doubly stochastic or bistochastic matrix is a matrix that is column
or left stochastic as well as row or right stochastic, meaning that
each of its rows likewise sums to unity: $\sum^{N}_{j=1}\stochasticmatrix_{ij}=1$.
Summing across each of the rows and then adding up the results must
yield $M$, whereas summing down each of the columns and then adding
up the results must yield $N$, so the commutativity of addition implies
that $M=\sum^{M}_{i=1}\sum^{N}_{j=1}\stochasticmatrix_{ij}=N$. Doubly
stochastic matrices are therefore always square matrices.

A permutation matrix is a doubly stochastic matrix whose entries are
all either $0$s or $1$s, with exactly a single $1$ in each row
and in each column. By Birkhoff's theorem (Birkhoff 1946)\nocite{Birkhoff:1946toseal},
every doubly stochastic matrix is expressible as a convex combination
of permutation matrices, meaning a linear combination with non-negative
coefficients that sum to $1$. Birkhoff's theorem therefore establishes
that all doubly stochastic matrices can be obtained or derived from
a probability-like combination of permutation matrices, where permutation
matrices, again, are a special case of doubly stochastic matrices.

One might wonder whether all stochastic matrices, even if they are
not doubly stochastic or even square matrices, can be derived from
matrices that are a special case of stochastic matrices. The major
claim of this paper is that the answer is yes, and that all stochastic
matrices can be derived from unistochastic matrices. A unistochastic
matrix $\matf{\stochasticmatrix}$ is a square matrix whose individual
entries are each the modulus square of the corresponding entry of
a unitary matrix $\matf U$, meaning that $\stochasticmatrix_{ij}={\vert U_{ij}\vert}^{2}$
(Stueckelberg 1952; Horn 1954; Thompson 1989; Nylen, Tam, Uhlig 1993;
{\.Z}yczkowski et al. 2003; Bengtsson 2004)\nocite{Stueckelberg:1952theuds,Horn:1954dsmatdoarm,Thompson:1989uln,NylenTamUhlig:1993oteopsonhasm,ZyczkowskiKusSlomczynskiWojciechSommers:2003rum,Bengtsson:2004tiobu}. It follows from the unitarity of the matrix $\matf U$ that a unistochastic
matrix $\matf{\stochasticmatrix}$ is always doubly stochastic, and
is thus, in particular, stochastic.

A corollary to the Born Representation Theorem shows that if the POVM
in the theorem forms an $N$-member PVM of its own, then the stochastic
matrix $\matf{\stochasticmatrix}$ in the theorem must be unistochastic.
The second theorem proved in this paper, herein called the Unistochastic
Theorem, then shows that even beyond this particular circumstance,
any stochastic matrix can be derived from a unistochastic matrix of
a bounded size by a suitable form of marginalization. This second
theorem therefore establishes a kind of primacy of unistochastic matrices
over all stochastic matrices, with potential applications to quantum
theory.

\section{Mathematical Preliminaries\label{sec:Mathematical-Preliminaries}}

\paragraph{Definition 2.1 (Stochastic Matrices): }

For $M$ and $N$ positive integers, an $M\times N$ stochastic matrix
$\matgreek{\stochasticmatrix}$ consists of real-valued entries $\stochasticmatrix_{11},\stochasticmatrix_{12},\dots,\stochasticmatrix_{MN}$
satisfying two conditions: non-negativity, 
\begin{equation}
\stochasticmatrix_{ij}\geq0,\label{eq:ColumnStochasticMatrixNonnegEntries}
\end{equation}
 and normalization, 
\begin{equation}
\sum^{M}_{i=1}\stochasticmatrix_{ij}=1.\label{eq:ColumnStochasticMatrixColumnsSumToOne}
\end{equation}
 In some references, stochastic matrices are said to be column or
left stochastic. From a numerical point of view, the entries of an
$M\times N$ stochastic matrix can serve as conditional probabilities:
\begin{equation}
p\left(i\given j\right)\equiv p\left(i\textrm{ given }j\right)\defeq\stochasticmatrix_{ij}.\label{eq:ConditionalProbabilitiesAsStochasticMatrixEntries}
\end{equation}

\paragraph{Definition 2.2 (Unistochastic Matrices): }

For $N$ a positive integer, a square $N\times N$ matrix $\matgreek{\stochasticmatrix}$
is said to be unistochastic if each of its entries $\stochasticmatrix_{ij}$
is the modulus square $\verts{U_{ij}}^{2}$ of the corresponding entry
$U_{ij}$ of an $N\times N$ unitary matrix $\matf U$ with entries
taken from the complex numbers, 
\begin{equation}
\matf U^{\adj}\matf U=\matf U\matf U^{\adj}=\matf 1,\label{eq:DefUnitarity}
\end{equation}
 where $\matf 1$ is the $N\times N$ identity matrix, 
\begin{equation}
\matf 1\defeq\diag{1,\dots,1}\defeq\begin{pmatrix}1 & 0\\
0 & \ddots\\
 &  & 1
\end{pmatrix}.\label{eq:DefIdentityMatrix}
\end{equation}
 That is, 
\begin{equation}
\stochasticmatrix_{ij}=\verts{U_{ij}}^{2}.\label{eq:DefUnistochasticMatrix}
\end{equation}
Every unistochastic matrix is, in particular, a (square) stochastic
matrix and is also row stochastic, meaning that each of its rows likewise
sums to $1$. These properties follow from two simple calculations:
\begin{equation}
\sum^{N}_{i=1}\stochasticmatrix_{ij}=\sum^{N}_{i=1}\verts{U_{ij}}^{2}=\sum^{N}_{i=1}\parens{\matf U^{\dagger}}_{ji}\parens{\matf U}_{ij}=\parens{\matf 1}_{jj}=\delta_{jj}=1,\label{eq:UnistochasticImpliesColumnStochastic}
\end{equation}
\begin{equation}
\sum^{N}_{j=1}\stochasticmatrix_{ij}=\sum^{N}_{j=1}\verts{U_{ij}}^{2}=\sum^{N}_{j=1}\parens{\matf U}_{ij}\parens{\matf U^{\dagger}}_{ji}=\parens{\matf 1}_{ii}=\delta_{ii}=1,\label{eq:UnistochasticImpliesRowStochastic}
\end{equation}
 where $\delta_{ij}$ is the usual Kronecker delta, 
\begin{equation}
\delta_{ij}\defeq\begin{cases}
1 & \textrm{for }i=j,\\
0 & \textrm{for }i\ne j.
\end{cases}\label{eq:DefKroneckerDelta}
\end{equation}
 A matrix that is both column stochastic and row stochastic is said
to be doubly stochastic or bistochastic. Hence, in particular, every
unistochastic matrix is doubly stochastic.

\paragraph{Definition 2.3 (Projection-Valued Measures or PVMs): }

For $N$ a positive integer, an $N$-member set of $N\times N$ matrices
$\matf{\projector}_{1},\dots,\matf{\projector}_{N}$ with complex-valued
entries are said to comprise a projection-valued measure (PVM) if
and only if they are self-adjoint, 
\begin{equation}
\matf{\projector}^{\adj}_{i}=\matf{\projector}_{i},\label{eq:PVMSelfAdjoint}
\end{equation}
 idempotent, 
\begin{equation}
\matf{\projector}^{2}_{i}=\matf{\projector}_{i},\label{eq:PVMIdempotent}
\end{equation}
 mutually exclusive or orthogonal (which contains idempotence as the
special case $i=j$), 
\begin{equation}
\matf{\projector}_{i}\matf{\projector}_{j}=\delta_{ij}\matf{\projector}_{i},\label{eq:PVMMutuallyExclusive}
\end{equation}
 complete or exhaustive, 
\begin{equation}
\sum^{N}_{i=1}\matf{\projector}_{i}=\matf 1,\label{eq:PVMCompleteness}
\end{equation}
 and have unit trace, 
\begin{equation}
\tr\left(\matf{\projector}_{i}\right)=1.\label{eq:PVMUnitTrace}
\end{equation}
 In particular, in being self-adjoint \eqref{eq:PVMSelfAdjoint} and
idempotent \eqref{eq:PVMIdempotent}, each such matrix $\matf{\projector}_{i}$
is a projection matrix and is therefore positive semidefinite: 
\begin{equation}
\matf{\projector}_{i}\geq0.\label{eq:PVMPositiveDefinite}
\end{equation}
 Note that not all of the foregoing conditions are logically independent.

\paragraph*{Definition 2.4 (Positive-Operator-Valued Measures or POVMS): }

For $M$ and $N$ positive integers, an $M$-member set of $N\times N$
matrices $\matf E_{1},\dots,\matf E_{M}$ with complex-valued entries
are said to comprise a positive-operator-valued measure (POVM) if
and only if they are positive semidefinite, 
\begin{equation}
\matf E_{i}\geq0,\label{eq:POVMPosSemidef}
\end{equation}
 and complete or exhaustive, 
\begin{equation}
\sum^{M}_{i=1}\matf E_{i}=\matf 1.\label{eq:POVMCompleteness}
\end{equation}
 Note that positive semidefiniteness \eqref{eq:POVMPosSemidef} implies
self-adjointness, 
\begin{equation}
\matf E^{\adj}_{i}=\matf E_{i}.\label{eq:POVMSelfAdjoint}
\end{equation}
 A POVM therefore generalizes the notion of a PVM. In particular,
a POVM need not have $M=N$ members in total ($M$ could be greater
than or less than $N$), and its members need not be idempotent, nor
must they satisfy a condition of mutual exclusivity or orthogonality
akin to \eqref{eq:PVMMutuallyExclusive}.

\section{Theorems\label{sec:Theorems}}

\paragraph{Theorem 3.1 (The Born Representation Theorem): }

For positive integers $M$ and $N$, let $\matgreek{\stochasticmatrix}$
be an $M\times N$ stochastic matrix with individual entries $\stochasticmatrix_{11},\stochasticmatrix_{12},\dots,\stochasticmatrix_{MN}$.
Then there exist
\begin{itemize}
\item an $M$-member positive-operator-valued measure (POVM) consisting
of $N\times N$ matrices $\matf E_{1},\dots,\matf E_{M}$ and 
\item an $N$-member projection-valued measure (PVM) consisting of $N\times N$
projection matrices $\matf{\projector}_{1},\dots,\matf{\projector}_{N}$ 
\end{itemize}
such that, in terms of a trace over $N\times N$ matrices, 
\begin{equation}
\keyeq{\stochasticmatrix_{ij}=\tr\left(\matf E_{i}\matf{\projector}_{j}\right).}\label{eq:PVM-POVMTheorem}
\end{equation}
 The reader familiar with quantum theory may recognize a resemblance
between this formula and a highly general form of the Born rule, which
gives the conditional probability $p\left(i\given j\right)$ for a
quantum system described by a pure state, as represented by $\matf{\projector}_{j}$,
to yield a given result or effect, as represented by $\matf E_{i}$,
when undergoing a generalized measurement.

As an immediate consequence, if each POVM matrix $\matf E_{i}$ has
unit trace, $\tr\left(\matf E_{i}\right)=1$, then $\matgreek{\stochasticmatrix}$
is doubly stochastic, as follows directly from the completeness relation
\eqref{eq:PVMCompleteness} satisfied by the PVM matrices $\matf{\projector}_{1},\dots,\matf{\projector}_{N}$.

\paragraph{Proof:}

Let $\matf{\projector}_{1},\dots,\matf{\projector}_{N}$ be any $N$-member
PVM of $N\times N$ matrices, such as, for example, the diagonal PVM
defined by 
\begin{equation}
\matf{\projector}_{i}\defeq\diag{0,\dots,0,1\ \left[i\textrm{th entry}\right],0,\dots,0},\label{eq:DefDiagPVM}
\end{equation}
 or, equivalently, in terms of individual entries, 
\begin{equation}
\projector_{i,jk}\defeq\delta_{ij}\delta_{ik}=\begin{cases}
1 & \textrm{for }j=k=i,\\
0 & \textrm{otherwise}.
\end{cases}\label{eq:DefDiagPVMComponents}
\end{equation}
 Let $\matf E_{1},\dots,\matf E_{M}$ be a POVM consisting of diagonal
matrices defined by the convex sums 
\begin{equation}
\matf E_{i}\defeq\sum^{N}_{j=1}\stochasticmatrix_{ij}\matf{\projector}_{j},\label{eq:DefDiagonalPOVM}
\end{equation}
 which are positive semidefinite \eqref{eq:POVMPosSemidef} on account
of the positive semidefiniteness \eqref{eq:PVMPositiveDefinite} of
the PVM $\matf{\projector}_{1},\dots,\matf{\projector}_{N}$ together
with the non-negativity of the coefficients, and are complete \eqref{eq:POVMCompleteness}
due to the normalization condition \eqref{eq:ColumnStochasticMatrixColumnsSumToOne}
on $\matf{\stochasticmatrix}$ together with the completeness \eqref{eq:PVMCompleteness}
of the PVM. It follows from the mutual exclusivity \eqref{eq:PVMMutuallyExclusive}
of the PVM together with their unit-trace condition \eqref{eq:PVMUnitTrace}
that 
\begin{equation}
\stochasticmatrix_{ij}=\tr\left(\matf E_{i}\matf{\projector}_{j}\right).\qquad\mathrm{QED}\label{eq:FinalFormulaPVM-POVMTheorem}
\end{equation}

\paragraph{Corollary 3.2:}

Given the assumptions and conclusion of Theorem~3.1, suppose that
the POVM matrices $\matf E_{1},\dots,\matf E_{M}$ form an $N$-member
PVM $\matf{\projector}^{\prime}_{1},\dots,\matf{\projector}^{\prime}_{N}$
of their own, with $\matf E_{i}=\matf{\projector}^{\prime}_{i}$ for
each $i=1,\dots,N$. Then 
\begin{equation}
\stochasticmatrix_{ij}=\tr\left(\matf{\projector}^{\prime}_{i}\matf{\projector}_{j}\right).\label{eq:ColumnStochasticMatrixAsTraceProductPVMs}
\end{equation}

\begin{itemize}
\item Because the $N$ members $\matf{\projector}_{1},\dots,\matf{\projector}_{N}$
of the first PVM are all self-adjoint projection matrices with unit
trace, they each have a single nonzero eigenvalue equal to $1$, and
so there exists an orthonormal basis $\matgreek{\epsilon}_{1},\dots,\matgreek{\epsilon}_{N}$
of mutual eigenvectors shared by the first PVM $\matf{\projector}_{1},\dots,\matf{\projector}_{N}$,
with the eigenvalue equations 
\begin{equation}
\matf{\projector}_{i}\matgreek{\epsilon}_{j}=\delta_{ij}\matgreek{\epsilon}_{j}\label{eq:FirstPVMEigenvalueEquation}
\end{equation}
 and the outer-product factorizations 
\begin{equation}
\matf{\projector}_{i}=\matgreek{\epsilon}_{i}\matgreek{\epsilon}^{\adj}_{i}.\label{eq:FirstPVMFactorizationBasisVector}
\end{equation}
\item Similarly, let $\matgreek{\epsilon}^{\prime}_{1},\dots,\matgreek{\epsilon}^{\prime}_{N}$
denote an orthonormal basis of mutual eigenvectors shared by the second
PVM $\matf{\projector}^{\prime}_{1},\dots,\matf{\projector}^{\prime}_{N}$,
with the eigenvalue equations 
\begin{equation}
\matf{\projector}^{\prime}_{i}\matgreek{\epsilon}^{\prime}_{j}=\delta_{ij}\matgreek{\epsilon}^{\prime}_{j}\label{eq:SecondPVMEigenvalueEquation}
\end{equation}
 and the outer-product factorizations  
\begin{equation}
\matf{\projector}^{\prime}_{i}=\matgreek{\epsilon}^{\prime}_{i}\matgreek{\epsilon}^{\prime\adj}_{i}.\label{eq:SecondPVMFactorizationBasisVector}
\end{equation}
\item Define an $N\times N$ matrix 
\begin{equation}
\matf V\defeq\sum^{N}_{i=1}\matgreek{\epsilon}_{i}\matgreek{\epsilon}^{\prime\adj}_{i}.\label{eq:DefUnitaryTransformationMatrix}
\end{equation}
 
\end{itemize}
It then follows that $\matf V$ is unitary, 
\begin{equation}
\matf V^{\adj}=\matf V^{-1},\label{eq:UnitarityTransformationMatrix}
\end{equation}
 and also that it relates the two PVMs according to 
\begin{equation}
\matf{\projector}^{\prime}_{i}=\matf V^{\adj}\matf{\projector}_{i}\matf V.\label{eq:UnitaryTransformationTwoPVMs}
\end{equation}
 Further, in terms of the individual entries $V_{ij}=\matgreek{\epsilon}^{\adj}_{i}\matf V\matgreek{\epsilon}_{j}$
of the matrix $\matf V$ with respect to the orthonormal basis $\matgreek{\epsilon}_{1},\dots,\matgreek{\epsilon}_{N}$,
it follows that 
\begin{equation}
\keyeq{\stochasticmatrix_{ij}=\verts{V_{ij}}^{2},}\label{eq:ColumnStochasticMatrixIsUnistochastic}
\end{equation}
 so $\matgreek{\stochasticmatrix}$ is a unistochastic matrix.

\paragraph{Proof:}

The unitarity \eqref{eq:UnitarityTransformationMatrix} of $\matf V$
and the transformation identity \eqref{eq:UnitaryTransformationTwoPVMs}
both follow from straightforward calculations, using the orthogonality
and unit-norm of the basis vectors. Then 
\begin{align*}
\stochasticmatrix_{ij} & =\tr\left(\matf{\projector}^{\prime}_{i}\matf{\projector}_{j}\right)=\tr\parens{\matf V^{\adj}\matf{\projector}_{i}\matf V\matf{\projector}_{j}}\\
 & =\tr\parens{\matf V^{\adj}\matgreek{\epsilon}_{i}\matgreek{\epsilon}^{\adj}_{i}\matf V\matgreek{\epsilon}_{j}\matgreek{\epsilon}^{\adj}_{j}}\\
 & =\parens{\matgreek{\epsilon}^{\adj}_{j}\matf V^{\adj}\matgreek{\epsilon}_{i}}\parens{\matgreek{\epsilon}^{\adj}_{i}\matf V\matgreek{\epsilon}_{j}}\\
 & =\parens{\matf V^{\adj}}_{ji}\parens{\matf V}_{ij}=\verts{V_{ij}}^{2}.\qquad\mathrm{QED}
\end{align*}

\paragraph{Theorem 3.3 (The Unistochastic Theorem):}

For $M$ and $L$ positive integers, let $\matgreek{\stochasticmatrix}$
be an $M\times L$ stochastic matrix with individual entries $\stochasticmatrix_{11},\stochasticmatrix_{12},\dots,\stochasticmatrix_{ML}$.
Note that $\matgreek{\stochasticmatrix}$ here is not assumed to be
unistochastic. Then for some positive integer $N$ equal to the greater
of $M$ and $L$, there exist
\begin{itemize}
\item an $N$-member projection-valued measure (PVM) consisting of $N\times N$
projection matrices $\matf{\projector}_{1},\dots,\matf{\projector}_{N}$, 
\item an $N^{2}$-member PVM consisting of $N^{2}\times N^{2}$ projection
matrices $\mathbb{\projector}_{\left(11\right)},\mathbb{\projector}_{\left(12\right)},\dots,\mathbb{\projector}_{\left(NN\right)}$, 
\item an $N^{2}\times N^{2}$ unitary matrix $\mathbb{U}$, and 
\item another $N^{2}$-member PVM consisting of $N^{2}\times N^{2}$ projection
matrices $\mathbb{\projector}^{\prime}_{\left(11\right)},\mathbb{\projector}^{\prime}_{\left(12\right)},\dots,\mathbb{\projector}^{\prime}_{\left(NN\right)}$
defined according to 
\begin{equation}
\mathbb{\projector}^{\prime}_{\left(ij\right)}\defeq\mathbb{U}^{\adj}\mathbb{\projector}_{\left(ij\right)}\mathbb{U}\label{eq:DefTransformedDilatedPVM}
\end{equation}
\end{itemize}
such that, in terms of a trace over $N^{2}\times N^{2}$ matrices,
the quantities 
\begin{equation}
\keyeq{\left(\stochasticmatrix\negthickspace\!\stochasticmatrix\right)_{\left(ik\right),\left(jl\right)}=\operator{Tr}\parens{\mathbb{\projector}^{\prime}_{\left(ik\right)}\mathbb{\projector}_{\left(jl\right)}}=\verts{\left(\mathbb{U}\right)_{\left(ik\right),\left(jl\right)}}^{2}}\label{eq:DilatedPVMTheoremUnistochasticMatrix}
\end{equation}
 are the entries of an $N^{2}\times N^{2}$ unistochastic matrix $\stochasticmatrix\negthickspace\!\stochasticmatrix$
with row multi-index $\left(ik\right)$ and column multi-index $\left(jl\right)$.
Moreover, the entries of the original $M\times L$ stochastic matrix
$\matgreek{\stochasticmatrix}$ can be expressed as the marginalization
formula 
\begin{equation}
\keyeq{\stochasticmatrix_{ij}=\sum^{N}_{k=1}\left(\stochasticmatrix\negthickspace\!\stochasticmatrix\right)_{\left(ik\right),\left(j1\right)}.}\label{eq:DilatePVMTheoremAsMarginalization}
\end{equation}

As an immediate consequence, the entries of the original $M\times L$
stochastic matrix $\matgreek{\stochasticmatrix}$ are expressible
as 
\begin{equation}
\keyeq{\stochasticmatrix_{ij}=\operator{Tr}\left(\mathbb{\projector}^{\prime}_{i}\mathbb{\projector}_{\left(j1\right)}\right),}\label{eq:DilatedPVMTheoremAsTrace}
\end{equation}
 where 
\begin{equation}
\mathbb{\projector}^{\prime}_{i}\defeq\sum^{N}_{k=1}\mathbb{\projector}^{\prime}_{\left(ik\right)}.\label{eq:DefTransformedDilatedTruncatedPVM}
\end{equation}

\paragraph{Proof:}

In principle, one can prove this theorem by using Theorem 3.1 and
Corollary 3.2 together with an appropriate version of the Naimark
dilation theorem (Naimark 1943)\nocite{Naimark:1943oaroaosf} or
the Stinespring dilation theorem (Stinespring 1955)\nocite{Stinespring:1955pfoc}.\footnote{The author thanks Gopalkrishnan (2026)\nocite{Gopalkrishnan:2026ispctqdofcs}
for tightening the bound on the dilation dimension in the original
version of this theorem from $N^{3}$ to $N^{2}$, as compared with
an earlier result (Barandes 2023, 2025)\nocite{Barandes:2023tsqt,Barandes:2025tsqc}.
Independently, Schmidt (2021)\nocite{Schmidt:2021dosmbcg} presented
a similar theorem, but the proof there appears to have an error in
its construction of the required partial isometry analogous to \eqref{eq:DefIntermedPartialIsometryMatrix}
in the present work.} To make the present treatment self-contained, a constructive proof
will now follow instead.

To begin, if $M=L$, then define $\tilde{\matgreek{\stochasticmatrix}}\defeq\matgreek{\stochasticmatrix}$,
and then set $N\defeq M$. Otherwise, if $M>L$, then extend $\matgreek{\stochasticmatrix}$
to an $M\times M$ stochastic matrix $\tilde{\matgreek{\stochasticmatrix}}$
by adding $M-L$ additional columns, where each new column contains
a single entry equal to $1$ and all its other entries are equal to
$0$, and then set $N\defeq M$. If instead $M<L$, then extend $\matgreek{\stochasticmatrix}$
to an $L\times L$ stochastic matrix $\tilde{\matgreek{\stochasticmatrix}}$
by adding $L-M$ additional rows, where each row contains only $0$s,
and then set $N\defeq L$. In any of these cases, the result is that
$\matgreek{\stochasticmatrix}$ is now embedded inside a square $N\times N$
stochastic matrix $\tilde{\matgreek{\stochasticmatrix}}$ in the sense
that 
\begin{equation}
\stochasticmatrix_{ij}=\tilde{\stochasticmatrix}_{ij}\label{eq:OriginalStochasticMatrixEmbeddingIntoExtended}
\end{equation}
 for all $i=1,\dots,M\leq N$ and $j=1,\dots,L\leq N$.

Invoking the non-negativity condition $\tilde{\stochasticmatrix}_{ij}\geq0$
from \eqref{eq:ColumnStochasticMatrixNonnegEntries}, let $\matgreek{\dynop}$
be any (non-unique) $N\times N$ matrix with real-valued or complex-valued
entries satisfying 
\begin{equation}
\tilde{\stochasticmatrix}_{ij}=\verts{\dynop_{ij}}^{2}.\label{eq:StochasticMatrixAsModSquareDynOp}
\end{equation}
 Equivalently, in terms of the diagonal $N\times N$ PVM matrices
$\matf{\projector}_{1},\dots,\matf{\projector}_{N}$ defined in \eqref{eq:DefDiagPVM},
together with a trace over $N\times N$ matrices, 
\begin{equation}
\tilde{\stochasticmatrix}_{ij}=\tr\parens{\matgreek{\dynop}^{\adj}\matf{\projector}_{i}\matgreek{\dynop}\matf{\projector}_{j}}.\label{eq:StochasticQuantumDictionary}
\end{equation}
 Then by the normalization condition \eqref{eq:ColumnStochasticMatrixColumnsSumToOne}
on the stochastic matrix $\tilde{\matgreek{\stochasticmatrix}}$,
the new matrix $\matgreek{\dynop}$ satisfies the sum rule 
\begin{equation}
\parens{\matgreek{\dynop}^{\adj}\matgreek{\dynop}}_{jj}=\sum^{N}_{i=1}\verts{\dynop_{ij}}^{2}=1.\label{eq:DynOpSumRule}
\end{equation}
 Define an $N$-member collection of $N\times N$ matrices $\matf{\krausmatrix}_{1},\dots,\matf{\krausmatrix}_{N}$
according to 
\begin{equation}
\matf{\krausmatrix}_{i}\defeq\matgreek{\dynop}\matf{\projector}_{i},\label{eq:DefKrausMatrices}
\end{equation}
 or, equivalently, in terms of components, as 
\begin{equation}
\krausmatrix_{i,jk}\defeq\dynop_{ji}\delta_{ik}.\label{eq:DefKrausMatricesComponents}
\end{equation}
 Then it follows from the sum rule \eqref{eq:DynOpSumRule} together
with a straightforward calculation that these new matrices provide
a factorization of the $N\times N$ PVM matrices $\matf{\projector}_{1},\dots,\matf{\projector}_{N}$
according to 
\begin{equation}
\matf{\projector}_{i}=\matf{\krausmatrix}^{\adj}_{i}\matf{\krausmatrix}_{i}.\label{eq:ProductAdjointKrausKrausEqProjector}
\end{equation}
  They also obey the useful identity 
\begin{equation}
\matf{\krausmatrix}^{\adj}_{j}\matf{\projector}_{i}\matf{\krausmatrix}_{j}=\tilde{\stochasticmatrix}_{ij}\matf{\projector}_{j}=\verts{\Theta_{ij}}^{2}\matf{\projector}_{j}.\label{eq:DoubleKrausMatrixProjectorIdentity}
\end{equation}
  The matrices $\matf{\krausmatrix}_{1},\dots,\matf{\krausmatrix}_{N}$
are therefore Kraus matrices (Kraus 1971)\nocite{Kraus:1971gscqt},
because, from the factorization \eqref{eq:ProductAdjointKrausKrausEqProjector}
together with the completeness relation \eqref{eq:PVMCompleteness},
they satisfy the Kraus identity, 
\begin{equation}
\sum^{N}_{i=1}\matf{\krausmatrix}^{\adj}_{i}\matf{\krausmatrix}_{i}=\matf 1,\label{eq:KrausIdentity}
\end{equation}
 and it follows from the identity \eqref{eq:DoubleKrausMatrixProjectorIdentity}
and another straightforward calculation that they give a Kraus decomposition
of the entries of the stochastic matrix $\tilde{\matgreek{\stochasticmatrix}}$,
\begin{equation}
\tilde{\stochasticmatrix}_{ij}=\sum^{N}_{k=1}\tr\parens{\matf{\krausmatrix}^{\adj}_{k}\matf{\projector}_{i}\matf{\krausmatrix}_{k}\matf{\projector}_{j}}=\sum^{N}_{k=1}\verts{\krausmatrix_{k,ij}}^{2}.\label{eq:StochasticMatrixKrausDecomposition}
\end{equation}
  Defining a new $N$-member collection of $N\times N$ matrices
$\matf E_{1},\dots,\matf E_{N}$ according to 
\begin{equation}
\matf E_{i}\defeq\sum^{N}_{k=1}\matf{\krausmatrix}^{\adj}_{k}\matf{\projector}_{i}\matf{\krausmatrix}_{k},\label{eq:DefPOVMFromKraus}
\end{equation}
 these matrices $\matf E_{1},\dots,\matf E_{N}$ simplify to the same
POVM matrices as \eqref{eq:DefDiagonalPOVM}, though now with $M=N$:
\begin{equation}
\matf E_{i}=\sum^{N}_{j=1}\tilde{\stochasticmatrix}_{ij}\matf{\projector}_{j}.\label{eq:POVMSimplifiedDiagonal}
\end{equation}

Next, letting $\left(ik\right)$ and $\left(jl\right)$ each denote
multi-indices running through $N^{2}$ values, define $N$ columns
of an $N^{2}\times N^{2}$ matrix $\mathbb{\timeevop}$ according
to 
\begin{equation}
\parens{\mathbb{\timeevop}}_{\left(ik\right),\left(j1\right)}\defeq\krausmatrix_{k,ij},\label{eq:DefIntermedPartialIsometryMatrix}
\end{equation}
 where the other columns will be defined shortly (Gopalkrishnan 2026)\nocite{Gopalkrishnan:2026ispctqdofcs}.
Then it follows from the Kraus identity \eqref{eq:KrausIdentity}
together with a straightforward computation that 
\begin{align}
\parens{\mathbb{\timeevop}^{\adj}\mathbb{\timeevop}}_{\left(m1\right),\left(j1\right)} & =\sum^{N}_{i=1}\sum^{N}_{k=1}\parens{\mathbb{\timeevop}^{\adj}}_{\left(m1\right),\left(ik\right)}\parens{\mathbb{\timeevop}}_{\left(ik\right),\left(j1\right)}=\sum^{N}_{i=1}\sum^{N}_{k=1}\overconj{\krausmatrix_{k,im}}\krausmatrix_{k,ij}=\sum^{N}_{k=1}\parens{\matf{\krausmatrix}^{\adj}_{k}\matf{\krausmatrix}_{k}}_{mj}=\left(\matf 1\right)_{mj}\nonumber \\
 & =\delta_{mj}.\label{eq:IntermedPartialIsometryMatrixProdEqId}
\end{align}
 Hence, the $N$ columns of $\mathbb{\timeevop}$ defined by \eqref{eq:DefIntermedPartialIsometryMatrix}
are each $N^{2}\times1$ matrices, 
\begin{equation}
\matgreek{\epsilon}_{\left(11\right)},\matgreek{\epsilon}_{\left(21\right)},\dots,\matgreek{\epsilon}_{\left(N1\right)},\label{eq:ColumnMatricesFromPartialIsometry}
\end{equation}
 and, due to \eqref{eq:IntermedPartialIsometryMatrixProdEqId}, are
mutually orthonormal. By the Gram-Schmidt process, they can be extended
to an $N^{2}$-member orthonormal basis $\matgreek{\epsilon}_{\left(11\right)},\matgreek{\epsilon}_{\left(12\right)},,\dots,\matgreek{\epsilon}_{\left(NN\right)}$.
The $N^{2}$ members of this orthonormal basis fully (though non-uniquely)
define all the columns of the $N^{2}\times N^{2}$ matrix $\mathbb{U}$
to make it into a unitary matrix, in the precise sense that 
\begin{equation}
\mathbb{U}^{\adj}\mathbb{U}=\mathbb{U}\mathbb{U}^{\adj}=\idmatrix_{N^{2}\times N^{2}},\label{eq:DefDilatedUnitary}
\end{equation}
 where $\idmatrix_{N^{2}\times N^{2}}$ is the $N^{2}\times N^{2}$
identity matrix.

Now using the diagonal $N\times N$ PVM $\matf{\projector}_{1},\dots,\matf{\projector}_{N}$
from \eqref{eq:DefDiagPVM}, introduce an $N^{2}$-member collection
of $N^{2}\times N^{2}$ projection matrices $\mathbb{\projector}_{\left(11\right)},\mathbb{\projector}_{\left(12\right)},\dots,\mathbb{\projector}_{\left(NN\right)}$
as the tensor products 
\begin{equation}
\mathbb{\projector}_{\left(ij\right)}\defeq\matf{\projector}_{i}\tensorprod\matf{\projector}_{j},\label{eq:DefDilatedPVMForProof}
\end{equation}
 meaning that 
\begin{equation}
\parens{\mathbb{\projector}_{\left(ij\right)}}_{\left(kl\right),\left(mn\right)}\defeq\left(\matf{\projector}_{i}\right)_{km}\left(\matf{\projector}_{j}\right)_{ln}=\delta_{ik}\delta_{im}\delta_{jl}\delta_{jn}.\label{eq:DefDilatedPVMEntries}
\end{equation}
Then define another $N^{2}$-member collection of $N^{2}\times N^{2}$
projection matrices $\mathbb{\projector}^{\prime}_{\left(11\right)},\mathbb{\projector}^{\prime}_{\left(12\right)},\dots,\mathbb{\projector}^{\prime}_{\left(NN\right)}$
according to 
\begin{equation}
\mathbb{\projector}^{\prime}_{\left(ij\right)}\defeq\mathbb{U}^{\adj}\mathbb{\projector}_{\left(ij\right)}\mathbb{U}.\label{eq:DefTransformedDilatedPVMForProof}
\end{equation}
 Finally, define an $N^{2}\times N^{2}$ unistochastic matrix $\stochasticmatrix\negthickspace\!\stochasticmatrix$
in terms of its individual entries by any of the equivalent formulas
\begin{equation}
\left.\begin{aligned}\parens{\stochasticmatrix\negthickspace\!\stochasticmatrix}_{\left(ik\right),\left(jl\right)} & \defeq\operator{Tr}\parens{\mathbb{\projector}^{\prime}_{\left(ik\right)}\mathbb{\projector}_{\left(jl\right)}}\\
 & =\operator{Tr}\parens{\mathbb{U}^{\adj}\mathbb{\projector}_{\left(ik\right)}\mathbb{U}\mathbb{\projector}_{\left(jl\right)}}\\
 & =\verts{\parens{\mathbb{U}}_{\left(ik\right),\left(jl\right)}}^{2}.
\end{aligned}
\qquad\right\} \label{eq:DefDilatedUnistochasticMatrixEntries}
\end{equation}
It then follows from the definition \eqref{eq:DefIntermedPartialIsometryMatrix}
and the Kraus decomposition \eqref{eq:StochasticMatrixKrausDecomposition},
together with a straightforward calculation, that 
\[
\sum^{N}_{k=1}\parens{\stochasticmatrix\negthickspace\!\stochasticmatrix}_{\left(ik\right),\left(j1\right)}=\sum^{N}_{k=1}\verts{\parens{\mathbb{U}}_{\left(ik\right),\left(j1\right)}}^{2}=\sum^{N}_{k=1}\verts{\krausmatrix_{k,ij}}^{2}=\tilde{\stochasticmatrix}_{ij},
\]
 where, by \eqref{eq:OriginalStochasticMatrixEmbeddingIntoExtended},
for $i=1,\dots,M$ and $j=1,\dots,L$, the equality $\tilde{\stochasticmatrix}_{ij}=\stochasticmatrix_{ij}$
holds.~$\textrm{QED}$ 

\section{Potential Applications\label{sec:Potential-Applications}}

As a potentially interesting set of applications, the Unistochastic
Theorem (Theorem 3.3) may have relevance to discrete-time, finite-state,
time-homogeneous Markov chains. These Markov chains include, as special
cases, deterministic processes that are logically reversible (or bilaterally
deterministic, in the language of Watanabe 1965)\nocite{Watanabe:1965cpip},
as well as logically irreversible deterministic processes that lose
information with time.

The dynamical behavior of each such process is fixed by a choice of
constant stochastic matrix $\matf{\stochasticmatrix}$. Letting $N$
be the finite number of states for such a process, and letting $\matf p\defeq\left(p_{1}\cdots p_{N}\right)^{\transp}$
be a contingent initial probability vector\textemdash perhaps a trivial
probability vector if the process begins with certainty in one of
its $N$ states\textemdash then subsequent probability vectors are
given after $n\in\mathbb{Z}^{+}$ discrete time steps by $\matf{\stochasticmatrix}^{n}\matf p$.
By the Unistochastic Theorem, such a process is associated (non-uniquely)
with a unitarily evolving quantum system on a Hilbert space of dimension
no greater than $N^{2}$, for some unitary matrix $\mathbb{\timeevop}$.

For example, consider a discrete-time, time-homogeneous, logically
reversible, deterministic process on a state space consisting of a
finite number $N$ of states. Then the stochastic matrix $\matf{\stochasticmatrix}$
consists only of $1$s and $0$s, with a single $1$ in each row and
in each column, so $\matf{\stochasticmatrix}=\matf{\Sigma}$ is an
$N\times N$ permutation matrix. Every permutation matrix is already
trivially a unitary matrix, and unitary matrices have well-defined
real-valued powers, so, for $t$ a real-valued time parameter, let
\begin{equation}
\matf{\timeevop}\left(t\right)\defeq\matf{\Sigma}^{t},\label{eq:DiscreteSystemUnitaryTimeEvolutionOp}
\end{equation}
 which defines a smoothly time-indexed family of unitary matrices
over the complex numbers. This time-evolution operator $\matf{\timeevop}\left(t\right)$
satisfies the requirements of Stone's theorem (Stone 1930)\nocite{Stone:1930ltihs},
so it has a generator $\matf H$, which can be identified as a quantum-theoretic
Hamiltonian. Letting $\vecgreek{\epsilon}_{k}$ denote an initial
basis vector corresponding to exactly one of the $N$ possible states,
one can define a complex-valued, time-evolving state vector $\vecgreek{\psi}\left(t\right)\defeq\matf{\timeevop}\left(t\right)\vecgreek{\epsilon}_{k}$,
and thus one obtains an analytic interpolation of the discrete-time
deterministic process as a continuous-time quantum process, with energy
eigenvalues and a Schrödinger equation. One can generalize to an initial
condition with a nontrivial probability distribution $\matf p\defeq\left(p_{1}\cdots p_{N}\right)^{\transp}$
by introducing an initially diagonal density matrix $\vec{\densitymatrix}_{0}\defeq\diag{p_{1},\dots,p_{N}}$
and then evolving it forward in time according to $\vecgreek{\densitymatrix}\left(t\right)\defeq\matf{\timeevop}\left(t\right)\vec{\densitymatrix}_{0}\matf{\timeevop}^{\adj}\left(t\right)$.

Now consider a discrete-time, time-homogeneous, logically irreversible,
deterministic process, such as a finite-state version of the Collatz
process defined for $k\in\left\{ 0,1,2,3,4\right\} $ by the dynamical
rule 
\begin{equation}
k\mapsto\begin{cases}
3k+1\textrm{ mod }5 & \textrm{for }k\textrm{ odd},\\
k/2 & \textrm{for }k\textrm{ even},
\end{cases}\label{eq:FiniteQuotientCollatzRule}
\end{equation}
 which translates to 
\begin{align}
0 & \mapsto0,\nonumber \\
1 & \mapsto4,\nonumber \\
2 & \mapsto1,\label{eq:FiniteQuotientCollatzRuleAllSteps}\\
3 & \mapsto2,\nonumber \\
4 & \mapsto2.\nonumber 
\end{align}
 This dynamical rule can be encoded into the $5\times5$ stochastic
matrix 
\begin{equation}
\matf{\stochasticmatrix}\defeq\begin{pmatrix}1 & 0 & 0 & 0 & 0\\
0 & 0 & 1 & 0 & 0\\
0 & 0 & 0 & 1 & 1\\
0 & 0 & 0 & 0 & 0\\
0 & 1 & 0 & 0 & 0
\end{pmatrix},\label{eq:FiniteQuotientCollatzStochasticMatrix}
\end{equation}
 which consists solely of $1$s and $0$s but is not a permutation
matrix because two of its columns are identical. One can, in principle,
write down a $25\times25$ unitary matrix $\mathbb{\timeevop}$ corresponding
to $\matf{\stochasticmatrix}$ in the sense of the Unistochastic Theorem,
and then introduce a Hamiltonian and a Schrödinger equation, as before.
However, unlike for a logically reversible process, as considered
in the previous paragraph, one would need to impose Hilbert-space
projections at each integer time step in order to capture the exact
behavior of the original deterministic process. Projections at integer
time steps would also generally be needed for Markov chains with genuinely
probabilistic dynamical rules.

\bibliographystyle{1_home_jacob_Documents_Work_My_Papers_2026-PVM-POVM_Theorems_custom-abbrvalphaurl}
\bibliography{0_home_jacob_Documents_Work_My_Papers_Bibliography_Global-Bibliography}

\end{document}